# Dirac's Book *The Principles Of Quantum Mechanics*

## as presenting the alternative arganization of a theory

Antonino Drago

formerly at Dept. Physical Sciences, University "Federico II" of Naples – Italy

**Abstract.** Authoritative appraisals qualified this book as an "axiomatic" theory. However, being its essential content no more than an analogy, its theoretical organization cannot be an axiomatic one. In fact, in the first edition Dirac declares to avoid an axiomatic presentation. Moreover, I show that the text is aimed at solving a basic problem (How quantum mechanics is similar to classical mechanics?). A previous paper analyzed all past theories of physics, chemistry and mathematics, presented by the respective authors in a non-axiomatic way. Four characteristic features of a new model of organizing a theory have been recognized. An accurate inspection of Dirac's text proves that it actually applied this kind of organization of a theory. This fact gives a formal reason of what Kronz and Lupher suggested through intuitive categories (pragmatism and rigor), i.e. Dirac's formulation of Quantum mechanics represents a distinct theoretical attitude from von Neumann's axiomatic attitude. However, since the second edition Dirac changed his attitude; although relying again on an analogy, his theory was referred to the axiomatic method. Some considerations on the odd paths which led to the present formulation of QM are added. They suggest to look for a new foundation of this theory.

**Keywords:** Dirac's book, Axiomatic organization, Problem-based organization, Mathematics' role in his book, Non-classical logic, New foundation

## 1. Which appraisals on Dirac's book?

In the historical development of Quantum Mechanics Paul A.M. Dirac's book *The principles of Quantum Mechanics* played a decisive role.(Dirac 1930) It illustrated his new theory of quantum mechanics (QM) for which he received the 1933 Nobel prize (together with Schroedinger).

A short and concise presentation of the book is the following one:

> **The Principles of Quantum Mechanics**. P. A. M. Dirac (Oxford U.P., London, I930, 1935, 1947, [1958, 1962, 2013]). This monograph is the first comprehensive text of the new quantum mechanics. and each new edition presents some important new view of the author. It is in a class by itself because it records the progress in his systematic development that starts from his insights into classical mechanics. The second edition presents the first approach to the path integral, and the third is phrased in the formalism of bra's < and ket's > whose juxtaposition forms the bracket < , > or scalar product. (A) (Gutzwiller 1998, p. 310 II)

Some appraisals by historians are the following ones:

> [It is] one of the classics of scientific thought. (D'Agostino 2002, p. 225)

> [The book] set[s] the stage, the tone, and much of the language of the quantum-mechanical revolution. (Brown 2006, p. 381)

> Dirac's formal framework for quantum mechanics was very useful and influential despite its lack of mathematical rigour… Working teachers as well as students of quantum mechanics often use Dirac's framework because its simplicity, elegance, power, and relative ease of use. Thus, from the standpoint of pragmatic, Dirac's framework is much referred over von Neumann's (Kronz and Lupher 2019, sect. 3).

The book is divided in two parts, the first dealing with a general physical and mathematical theory of QM, the second treating applications. I will confine myself to the first part.

The common reading recognizes in this part of Dirac's book a theory which is developed through abstract mathematical notions, to be interpreted in physical terms through a minimum of experimental facts.

About the entire Dirac's presentation, David Jammer maintained without argumentations that it presents the

> general formalism of quantum mechanics in a logically consistent and *axiomatic* fashion, based on the notion of "observable" and "state" as primitives. (Jammer 1989, p. 389; emphasis added).

Later, Helge Kragh offered two important reviews of the celebrated Dirac's "abstract textbook".(Kragh 1990, pp. 76ff.; Kragh 2013)[1] He presents it in a way that does not distinguish its organization from an axiomatic one (AO).

> The [symbolic] method was based on "certain symbols which we say denote physical things [… and which] we shall use in algebraic analysis in accordance with certain axioms" (I, p. 18)[2] (Kragh 2013, p. 261)

Here one can attribute the word "axioms" also to physical laws and principles, so that one can intend the book as a physical axiomatic. This interpretation is strengthened by one more Kragh's consideration.   Dirac qualified his brilliant mathematics, the algebra of a vector space and its group of transformations, as the "symbolic" method. Kragh illustrates it by recalling that Dirac stated that the

> "symbolic method […] deals directly in an abstract way with the quantities of fundamental importance (the invariants, &c., of the transformations)." This method, Dirac said, "seems to go more deeply into the nature of things." In accordance with the symbolic method, he wanted to present the general theory of quantum mechanics in a way that was free from physical interpretation [i.e. he wanted to avoid discussions on Copenhagen interpretation]: "One does not anywhere specify the exact nature of the symbols employed, nor is such specification at all necessary. They are used all the time in an abstract way, the algebraic axioms that they satisfy and the connexion between equations involving them and physical conditions being all that is required. [I, p. 18]" (Kragh 1990, p. 78)

The last propositions recall David Hilbert's celebrated presentation of a formal axiomatic; the meanings of the basic notions are given by the propositions deduced from the axioms. As a fact, most scholars have considered his new mathematical approach as a mere variant of an axiomatic theory. All in the above leads a reader to rally Dirac's attitude with Janos von Neumann's axiomatic one, which was heavily and rigorously based on Hilbert's space in an axiomatic way; (von Neumannn 1932) so much to assimilate Dirac's approach to subsequent von Neumann's approach. Therefore it is a custom to speak of "Dirac-von Neumann's foundation of QM".

But von Neumann (1932, Preface) remarked that Dirac's method "in no way satisfies the requirement of mathematical rigor." Recently, Kronz and Lupher (2019) stressed this divergence till up to consider Dirac's method for founding QM (and field theory too) in opposition to von Neumann's. These authors support their thesis by means of philosophical arguments based on the two intuitive notions, "rigor" and "pragmatism" (whose an instance is Dirac's very useful but ill-defined function $\delta(x)$).

A previous paper illustrated a comparative analysis of all scientific theories which have been presented by the respective authors in a non axiomatic way.(Drago 2012) The characteristic features of a new model of organization have been extracted; it is called problem-based (PO) because a PO theory is based on the search for a new method capable to solve a basic problem; it differs in a clear-cut way from AO, since AO is governed by classical logic whereas PO by intuitionist logic. Since he validity or not of the law of double negation makes the two kinds of logic mutually incompatible, there exists an antagonistic opposition between the two kinds of organization. This opposition constitutes a dichotomy in the foundations of TP.

In the following next sections 2-7 I will analyze the organization of the theory within Dirac's presentation of the first edition of this book. I will stress Dirac's declaration of his disbelief in the axiomatic method. Then I will show that the organization of the theory illustrated by the first edition of Dirac's book presents all the characteristic features of a PO. In sect. 8 I will analyze the great role played by mathematics in his book. Next sect. will suggest an appraisal of Dirac's theoretical attitude and its relevance for the further theoretical advancements of the formulation of QM. In sect. 10 I will present the important differences of the first editions with the next editions. Dirac radically changed attitude; only the axiomatic was considered by him as the possible organization of a theory. Through this change his attitude approached Janos von Neumann's attitude. In the conclusive sect. 11 I will remark that in the same year 1935 of the second edition of Dirac's book also von Neumann changed his basic mathematical attitude, from analytical attitude based on Hilbert's space to an algebraic one. These facts show the strange paths which have led to QM and suggest to look for a new formulation based on a new foundational basis.

---

[1]   Also Robert Oppenheimer underlined its abstractedness (quoted by Kragh 1990, p. 79 and Kragh 2013, p. 250); see some more appraisals in this page and the following one of latter Kragh's paper. Remarkably, a review of this book ended with the following words: "here is a physicist who is approaching the highest department of modern physics with so completely abstract point of view that he is regarded as a pure mathematician by many of his collegues."(Koopman 1935, p. 474)

[2]    A number of page inside round brackets refers to Dirac's book; the previous capital Roman number refers to the edition number.

**2. Book's first edition[3], Chapter 1: Dirac's disbelief in the axiomatic organization of a theory**
In order to present the new kind of organization of Dirac's theory let us start by a quick analysis of Chapter I and then Chapter VI; the preface and the mathematical development will be presented in next sections.

A basic philosophical question, which is independent of the kind of mathematics assumed by a theoretical physicist, concerns the the kind of organization of a theory. At the time of this book most scholars distinguished the theories formalized in an axiomatic way from non axiomatized theories, variously called as genetic, phenomenological, instrumental, empirical, etc.. About Dirac's organization of the theory Kragh (1990, p. 275) stated that "Dirac distinguished between the inductive-empirical and the mathematical-deductive method in science.", where the former one was barely considered a theoretical method and the latter one may be identified with the method of AO.[4] Bokulich (2008, chp. 3) attributes to Dirac the conception of a physical theory as an "open theory", undergoing a process of ever more improving its previous system; in any case a theory organized in a different way from the dominant organization of a theory. Unfortunately, these authors do not offer sharp criteria for distinguishing their alternatives to the axiomatic organization.

Let us investigate his book. In my opinion its title "*The Principles of Quantum Mechanics*" represented a sensational claim. In previous thirty years the certainties of theoretical physicists have been upset by quantum novelties; the subsequent search for some new certainties had been groping; instead Dirac's title echoed the title of Newton's celebrated book whose theory had governed theoretical physics along some centuries. In his time none could understand this title differently from a proclaim of having eventually achieved the very first principles of QM. Dirac's words seemed to announce a well-defined set of principles of an AO theory.

Author's claim seemed confirmed since the beginning of the book[5]; a "principle" is presented by first chapter's title ( "The superposition principle"); which is straight away dealt with in the first pages and then discussed till up the last section (no. 6). In the lack of a suggestion for a different interpretation of Dirac's word "principle", the reader thinks this word as an axiom, from which one then deduces a lot of consequences. This idea is also supported by an important Dirac's subsequent statement; he announces that even the most known and impressive principle of QM is derived from the superposition principle:

---

[3]   See the online edition: https://archive.org/details/in.ernet.dli.2015.177580/page/n27/mode/2up

[4]   Actually, already Poincaré (1903 Ch. "Optique et Electricité"; 1905, Ch. VII) had distinguished two kinds of physical theory. Yet, his characterization was largely ignored. The same occurred for Einstein's characterization of two organizations of a scientific theory, i.e. a "principle theory" (e.g. phenomenological thermodynamics) and "constructive theory" (e.g. statistical mechanics). (Frisch 2006) They correspond to respectively PO and AO (contrarily to the intuitive meanings of Einstein's words). Unfortunately, Einstein's appraisal on Dirac's book – "we owe to Dirac the most logically perfect presentation of this theory [QM]" (quoted in Kragh 2013, p. 251), corresponds to both kinds of organization, depending on the two ways of intending the adverb "logically" according to classical logic or intuitionist logic.

[5]   Notice that the titles of the sections and their sequence are not so much informative. The title of sect. 3 ought to include the word "state"; section 4 includes indeterminacy, a subject already treated by previous section; sect. 5 adds a "further discussion on photon", which actually introduces a qualification that "we shall find from [the new] mathematical theory". Similar remarks apply to the sections and their sequence of next chapters. Hence, his synthetic presentation of the book obliges a reader to discover by himself the real structure of the theory.

> The indeterminacy […] is a necessary consequence of the superposition relationship that quantum mechanics requires to exist between the states."(I, p. 10)

Actually, he presents his theoretical organization, yet in an odd way, since this subject occurs after a long illustration of many subjects and it is dealt with in the contrary direction to reader's first appraisal based on previous Dirac's words. In the last section (no. 6) he eventually achieves a definition of the principle announced by the title of the chapter. Just after, he closes the section and the entire chapter by discussing the organization of his theory. Here, he declares a different choice from mainstream's, by explicitly warning the reader to not interpret his theory as an axiomatic one.

> One could proceed to build up the theory of quantum mechanics on the basis of these ideas of superposition with the introduction of the minimum number of new assumptions necessary. Although this would be the logical line of development, it does not appear to be the most convenient one, as the laws of quantum mechanics are so closely interconnected that it would be not easy, and would be in any case somewhat artificial, to separate out the barest minimum of assumptions from which the rest could be deduced [….. Rather, my purpose is to] merely show the reasonableness and self-consistency of our fundamental assumptions. (I, pp. 16-17).

**3. Book's Chapter 1: Dirac's embarrassed justification of his non-axiomatic organization**

Both his expressions - "minimum number of new assumptions necessary" and "to separate out the barest minimum assumptions, from which the rest could be deduced" – apparently allude to the AO of a theory; but, strangely enough, he does not mention the word axiomatic.

However, from Dirac's words it is clear that he wants to avoid an axiomatic. He justifies his detachment from the commonly shared theoretical organization by means of three arguments. In a first time he states that this organization is not "convenient" because it is not easy to disentangle the "interconnections [of the laws]". Notice that this justification is weak; the mainstream attributed exactly this characteristic feature, "the interconnection of the laws", to a first stage (called "genetic", "phenomenological", "primitive", etc.) of the process of improving a theory till up to the unavoidable organization, AO.

After these apparently defensive words, Dirac expresses an aggressive criticism towards axiomatic method; it is "in any case somewhat artificial". Here, he manifests his distrust in axiomatic as a faithful representation of all physical theories. Yet, the mainstream usually rejected this criticism ("artificial") as coming from backwards theorists. This charge of backwardness well applies to Dirac's book, which at last concedes, in previous quotation, that his theory is based on, instead of some assured principles-axioms beginning a deductive system, a mere "reasonableness... of his fundamental assumptions".

As a third argument Dirac claims the "consistency of our fundamental assumptions". Yet, after Hilbert's program a consistency proof of a theory requires much more than some intuitive-heuristic argumentations as those offered by Dirac' following illustration (i.e. a recovering basic equations of QM, new results, successful applications, etc.).

In sum, he does not want assemble the theory as an AO. Rather, Dirac suggests his own "method" through the propositions that temporarily I missed in previous quotation:

> The method that will be here followed will therefore be first to give all the simple general laws in the form in which they are most easily expressed and remembered, and then to work out their consequences. This will mean that we shall continually be deducing results that are obviously necessary for the physical meaning of the theory to be tenable, or that follow from the foregoing ideas of superposition.

The "method" suggested by Dirac is not clear: in order to achieve an accurate description of a theoretical system, the words "simple", "easily expressed and remembered", "obviously necessary", "tenable" would have to be qualified better.[6]

---

[6] Let us recall that also the mathematicians opposing to Hilbert's axiomatic organization – i.e. Brouwer and partly Weyl and Heyting - never characterized how organize a theory in a new way, and maybe they even did not imagine that this new organization exists.

It is apparent that Dirac, being unsatisfied of the axiomatic method, chose a not well-defined alternative. That constitutes an embarrassed theoretical position. For this embarrassment he put in an unconventional location his warning about the organization of his theory; he postpones it at the end of chapter I.

As a matter of fact, the subsequent part of his theory attributes a central role to an analogy (between Poisson brackets (PB) and commutators).[7]

> Dirac had originally arrived at his formulation of quantum mechanics by noticing a close analogy between the Poisson brackets of classical dynamics and the non-commuting products found by Heisenberg, and he continued to find the analogy significant. As he showed in *Principles*, by means of the Poisson formulation of the classical equations of motion, "one can in this way obtain a quantum theory of individual dynamical systems analogous to the classical theory" (I, 93). The emphasis on the classical analogy was a special feature of Dirac's textbook and reflected his own discovery of quantum mechanics in the years 1925–26. (Kragh 2013, sect. 5, p. 261)

Yet, inside a theory an analogy manifestly obstructs an AO because, owing to its inaccurateness, it cannot play the role of an axiom and moreover it cannot be derived from any set of certain axioms. Moreover, in classical theoretical physics an analogy has been under-evaluated as little more than a guess. It is remarkable that Dirac is well-aware of the weakness of the central point of his theory; he openly admits that a mere analogy may be misleading:

> for example, the polariscope can be considered to be either a part of the system or a "disturbance." Similar arbitrariness is involved in the definitions of *preparation* and *observation*. Although superposition also occurs in classical wave theory, "*the superposition that occurs in quantum mechanics is of an essentially different nature from that occurring in the classical theory*. The analogies are therefore very misleading."(I, p. 11)

However at Dirac's time many physicists made use of analogies according to Bohr' correspondence principle between classical mechanics (CM) and the theory of quanta; in first decades of the century this "principle" has gained great importance because it suggests important results.[8] Dirac's mere analogy was a valuable theoretical suggestion only because at that time most quantum theorists applied (mainly through Bohr's correspondence principle) many analogies with classical mechanics. But Dirac's purpose to found *an entire theory* upon an analogy was however questionable.

Being well-aware that the great mathematicians, Hilbert and von Neumann, were very authoritative supporters of the axiomatic method, he had to defend his choice for a theoretical organization differing from the AO.

By suspecting hostile reactions to the unconventional organization of his theory, which in addition was founded on no more than an analogy, Dirac did not want to openly declare a sharp separation from the accredited organization; in the above quotation he mildly qualified as "*somewhat* artificial" (my emphasis) an axiomatic organization; he had to fear that an unrestricted rejection of it would lead most scholars to under evaluate, if not condemn as un-scientific his differently organized theory.[9]

In my opinion, it is for this reason that instead of locating the above quotation - which is a basic advice to a reader of the book - either within the "Preface" or at latest in the first rows of the first chapter, he put it at the end of this chapter, at the risk that the reader intends it as a merely additional note.

---

[7] He does not hide his appeal to an analogy. In the first edition of the book he makes use of this word in the title of sect. 33 and in several pages, e.g. pp. 11 (four times), 84, 86, 88, 93, 94, 95, 96, 98. 106.

[8] However Dirac's book never mentions Bohr's principle of correspondence (Mehra and Rechenberg 1986; vol. VI, p. 331). See also (Darrigol 1992, pp. 326-327).

[9] It seems not a case that mathematician Garrett Birkhoff's appraisal on this book stresses the typical requirement of a traditional scientist: "He impresses me as being at least comparatively deficient in appreciation of quantitative principles, logical consistency and completeness, and possibilities of systematic exposition and extension of a central theory."(quoted in Kragh 1990, p. 280).

In addition, in presenting his theory he had to be cautious, because Heisenberg had noticed an objection; this Dirac's analogy leads also to a contradiction (Mehra and Rechtenberg 1982, p. 331).

Moreover, he had to break the refractory attitude of most contemporary physicists to his mathematics of "transformations" (which is "really group theory"; Brown 2006, p. 386): let us recall Schroedinger's condemnation of the use of group theory within QM as a "*Gruppenpest*".[10] However, Dirac's presentation of group theory in his theory was singular. Kragh remarks that:

> Contrary to Weyl's textbook of 1928, which was based on the mathematical theory of groups, [a mention of] group theory was absent from *Principles* both in its first and later editions. Dirac was familiar with the new, mathematically abstract wa y of representing quantum theory, but he did not find it either more fundamental or very helpful. He preferred to treat group theory as part of quantum mechanics which for him was the general science of non-commutative quantities. (Dirac 1929)[11]

All in all, Dirac's above quoted words appear as a weak justification for his basic choice for a different organization of his theory from the authoritative AO. The great effort of his book is therefore to give as much as possible experimental and theoretical evidence for an analogy playing a central role within his theory.

## 4. How Dirac actually organized his theory? His strategy

After the Chapter I the reader is not advised about how long the march amid Dirac's new mathematics will be in order to see the goal of his effort of learning a lot of new notions. After Chapter I, reader's expectation for knowing more physical subjects of QM (and hence also more principles) has to wait the Chapter VI on "The equations of motion". The intermediate chapters ("II. Symbolic algebra of states and observables", "III. Eigenvalues and eigenstates". "IV. Representations of states and observables". "V. Transformation theory") constitute a mere mathematical preparation to formalize the physical content of Chapter VI ("Equations of motion and quantum conditions").

In this Chapter he declares the strategy of his previous illustration of the entire subject.
> The theory that has been developed so far contains a complete account of the new concepts and mathematical machinery required in quantum mechanics and also all the general physical laws. Only the general properties of states and observables have, however, been discussed, no reference being made to the particular conditions that they satisfy in the case of a specified dynamical system. (I, p. 92)[12]

It is not before this Chapter VI that the reader becomes aware that previous sophisticated formalism and previous claim – i.e. to have found out the basic principles of the new theory - result functional to present, in the lack of other more assured support, an analogy as the best framework for his theory. Here Dirac makes manifest that the core of his theory is an analogy between Poisson Brackets (PB) and commutators. This analogy compares a mathematical tool employed by classical Hamiltonian mechanics, PB, with a physical relationship of the theory of quanta: $pq - qp = ih/2\pi$ (where $p$ and $q$ are the state variables). This comparison makes apparent Dirac's strategy of his entire presentation. In previous Chapters Dirac had to accumulate as much as possible evidence for this assertion on both sides, the physical side and the mathematical one. On the former side, in Chapter I he introduces the notions of state, observable, probability, superposition principle, indeterminacy. On the mathematical side, he cumulates the notions of the chapters II-V. Then the aim of Chapter VI is to present (not prove!) the above analogy, constituting the central point of the theory.

---

[10] (Pauli 1932) Both Schroedinger and Slater were two annoyed physicists by Weyl's mathematical attitude. Weyl himself echoed the widespread criticisms that his book had received, when in the Preface of the second German edition (1931) he wrote: "It has been rumoured that the "group pest" is gradually being cut out of quantum physics"(p. x). See also Wigner (1959, p. v), that denounces a past, "great reluctance among physicists toward accepting group theoretical arguments and the group theoretical point of view"

[11] (Kragh 2013, sect. 6, p. 264) He suggested the following explanation: "Dirac was familiar with the new, mathematically abstract way of representing quantum theory, but he did not find it either more fundamental or very helpful. He preferred to treat group theory as part of quantum mechanics, which for him was the general science of non-commuting quantities." (Here Kragh paraphrases a quotation whose original text is reported by Darrigol 1992, p. 302, fn. 29). In my opinion this explanation does not gives reason why Dirac emphasised, in the beginnings of the "Preface", the theoretical novelty of "transformations" and moreover the basic role played by transformations for establishing the analogy. Moreover, Kragh's and Darrigol's quotation is of the year 1929, i.e. before the date of book's first edition.

[12] By incidence, it is not clear what above Dirac words "a complete account… of all general physical laws" mean, because no physical law is mentioned in previous chapters (apart what he called the superposition principle). Even the subsequent proposition seems deny previous claim of have covered "all general physical laws".

On the side of physics in Chapter I he suggests the basic notions. But even the basic words of QM are different from CM's. Hence hehas to define the vocabulary of QM needed for presenting the formulas of commutations. He starts from what is more shocking in QM vocabulary: the new notion of wave-corpuscle which joins in a peculiar way the two classical notions. Through this new notion he rightly stresses the radical change introduced by QM in the language of theoretical physics. After this cultural shock he leads the reader to look for an experimental basis of the theory. First, he introduces a new notion with respect to classical mechanics, a system state; it is defined as the situation of the measurement apparatus. In its turn, its presentation requires the description of the novelty of a physical experiment in QM, in particular a quantum jump. Dirac chooses the more simple as possible experiment, a polarization detection.

In this case the state of the system is defined by a two-valued variable (instead of the general case of two infinite valued variables, e.g. position and momentum). In this very simple case he shows that QM's notions are different, not the values , possibly discrete, of a state variable, but also the notion representing a totality, the entire system state. Owing to this demand, his definition of the latter is not easy in the quantum case; as a fact, Dirac presents it three times.

Through the above exemplary experiment more notions are verbally introduced: observable, probable value of an experiment result, indeterminacy (established in the year 1928 by Werner Heisenberg) to whose any emphasis is negated. In the last section (no. 6) he eventually defines the superposition principle. In sum, he suggests that this new basis is not an overcoming a barrier to our knowledge, but a new way of looking at physical experiments according to the superposition principle, which is a very novelty.

All in all, he presents the new basis of QM in an authoritative way. When describing an experiment on a system, nothing is proven in physical terms and mathematical either, all is merely told as a narrative to be assumed as an indisputable background. In addition, without argumentations, he asks reader to accept that the superposition principle "forms the fundamental new idea of quantum mechanics and the basis of the departure from the classical theory."(p. 2).

From here on, the next four chapters are devoted to translate all necessary physical notions into mathematical terms. Already Heisenberg formulation employed matrices, naturally suggested by commutators. Dirac goes to the origin of this mathematics by referring to vectors in a vector space.

In Chapter II a state of a physical system is translated into his basic mathematical notion, a vector, i.e. a mathematical notion which represents the superposition principle through a linear addition. As a consequence, all physical notions may be embedded into a vector space, equipped with its dual space (given a priori, without a physical justification).

## 5. The problem-based organization of Dirac's theory: the basic problem

Let us now skip to Chapter VI, whose first section deals again with the subject of the organization of his theory. This section surely represents a critical point of Dirac's strategy, also because its subject, the equations of motion, plays a basic role within a physical theory, as in Newton's mechanics the second principle, $F=ma$, does. In the following I will prove that Dirac's illustration of QM actually presents a PO theory.[13]

---

[13] He seems to allude to a choice for PO when he recalls his previous education. "I think that if l had not had this engineering training, I should not have had any success with the kind of work that I did later on, because it was really necessary to get away from the point of view one should deal only with results which could be deduced logically from known exact laws which one accepted, in which one had implicit faith."(quotes in Kragh 1990, p. 280) "Again in 1965,

As first step for recognizing a PO theory one has to inquire about which is the basic problem.

Surely, Dirac does not deals with the main philosophical problem that his contemporary physicists debated, i.e. Copenhagen interpretation of QM; he wants to present QM as free of any philosophical interpretation - as Kragh remarked (Kragh 1990, p. 78).[14] Rather, he confines himself to the new theoretical basis of QM.

Already in the first part of "Preface" Dirac stressed that classical mechanics is no longer an adequate theory with respect to quantum phenomena. Already the "Preface" stressed the cultural revolution represented by QM within the history of theoretical physics. Then the first page of Chapter I stresses a great theoretical *problem*, i.e. a ("irreconcilable") conflict between a classical description and quantum the experimental facts. Dirac recalls that in the first quarter of 20$^{th}$ century the passage from CM to QM had de-established the entire physical knowledge, owing to "The necessity for a fundamental departure from the laws and concepts of classical mechanics…".(I, p. 1)

Then, in the first page of Chp. VI he focuses the problem; he declares:

> If we are dealing with a given dynamical system, we shall have given dynamical variables, whose values at any time are what we call observables, and we shall require conditions that will determine the values of these variables at all times when their values at some particular time are known. These conditions will be the equations of motion of the system. In classical mechanics they would be sufficient to form the mathematical specification of the dynamical system under consideration. This is not so, however, in quantum mechanics.(I, p. 93)

Hence, Dirac is well aware that he was trying to solve *a great general problem*, that is how to change CM dynamics into QM dynamics.

In a first time his specific, mathematical problem is presented by declaring his goal:

*It is only when the quantum conditions are given as well as the equations of motion that we know as much about the variables as in the classical theory and can consider the dynamical system as mathematically completely specified.*(I p. 93)

Then he clarifies what are the quantum conditions:

> […] in quantum […] mechanics additional relations are necessary for this purpose, which take the form of equations connecting the values of the variables at a particular time, of such a nature that they can replace the commutative law of multiplication of the classical theory. These additional relations are called quantum conditions.(I p. 93)

Now the specific "problem" is stated in accurate terms:

> Our problem is now to determine the quantum conditions and [as a consequence] equations of motion for any given dynamical system, such as that formed by given electrons and atomic nuclei interacting. (I p. 93)

How then discover these "quantum conditions"?

> It is known that classical mechanics gives an accurate description of dynamical systems under certain limiting conditions, e.g. when the masses are large [with respect to the dimensions of the atomic particles]. One would therefore expect to be able to obtain a theory of these systems when the limiting conditions do not hold by making some natural generalizations in the classical equations of motion and choosing quantum conditions that form natural generalizations of the classical conditions that all the variables commute. [result:] It will be found that one can in this way obtain a quantum theory of individual dynamical systems *analogous* to the classical theory. (I p. 93, emphasis added)

Then Dirac announces that in the following he will answer this problem by means of a "natural generalization" of CM to "analogous dynamical systems" through which he will solve (almost) all quantum problems.

> It will be found that we can in this way obtain a quantum theory of individual and quantum systems analogous to the classical theory. ((1 p. 93).

---

he recommended that physicists… follow the more modest path of setting up "a theory with a reasonable practical standard of logic, rather like the way engineers work.""(quoted in Kragh 1990, p. 281) In the first quotation we see an implicit reference to his disbelief in an AO and in the second quotation his leaning to a PO.

[14] About Dirac's views on Copenhagen interpretation Kragh (1990, pp. 80-86) compares them with those of the most prominent physicists of his time. Dirac supported an independent view of mainstream's, but similar to it.

At this point, it is manifest that Dirac is presenting an at all different kind of organization from AO: first of all a problem, then a generalization aimed at eventually making plausible an analogy.

**6. The problem-based organization of Dirac's theory: propositions of non-classical logic**
More evidence for a PO theory is given by the occurrences of DNPs. They play an essential role within Dirac's illustration of his theory.[15]

Notice that an analogy may be translated into a doubly negated proposition; in our case "It is not true that quantum mechanics is not the same of classical mechanics", the corresponding affirmative proposition is not equivalent for the lack of evidence (DNP).[16]

Moreover, notice that the superposition principle is a DNP; it means neither that two states are equal, nor that they are distinct, but that each state is not severed from the other state. As a DNP, it surely makes impossible an AO of its theory. Like all DNPs beginning a PO organization, it represents rather than an axiom, a methodological principle; i.e. it plays the role of orienting the search for a new method capable to solve the problem stated at the beginning of the theory. This principle plays a similar role to that played by the impossibility of a perpetual (= without an end) motion within past theretical developments of mechanics and thermodynamics.

In addition, an analysis of Dirac's text shows that also his arguing relies on a such DNPs. For brevity' sake I do not record those occurring within the first chapter (there I recognized at least 15 DNPs). I only take into account those occurring at the beginnings of chp. VI, within the crucial section of the textbook, the no. 31, titled "The equations of motion and quantum Mechanics". (I, pp. 92-93)

1) We now must consider the form of these particular conditions and so make the theory applicable to given physical problems.... (I, p. 92)
2) ... we are concerned not with general physical laws applying to the whole (= not specific) of nature, but with special assumptions referring to a given specific physical problem... (I, p. 92)
3) Future developments of the theory may show that these assumptions are only approximate.(I, p. 92)
4) On the other hand, the assumptions of the four preceding chapters are so closely interconnected that one could hardly modify them in any way.... (I, p. 92)
5) ... without getting an entirely different scheme of mechanics,... (I, p. 92)
6) ... and the successes of the theory are so great as to make it fairly certain that no such modifications will be required.(I, p. 92)
7) The theory of these four chapters is in agreement with the principle of relativity; in fact it is so general to be in-dependent of any special relation between space and time... (I, p. 92)
8) ... We must, of course, for this to be true, adopt a more general definition of observable... (I, p. 92)
9) An observable now need no refer to an instant of time in some frame of reference...(I, p. 93)
10) ... so that there is no conflict with relativity on this account. (I, p. 93)
11) *It is only when quantum conditions are given as well as the equations of motion that we know as much about the variables as in the classical theory and can consider the dynamical system as mathematically completely specified...*(I, p. 93)
12) ... The equations of motion and quantum conditions are very closely connected with each other, and one cannot make any progress in solving a problem until they are [not] both solved.(I, p. 93)
13) When the limiting conditions do not hold (I, p. 93).

---

[15] In the following each negation of a DNP will be underlined in order to make easier the inspection by a reader; whereas each modal word or a single word which is equivalent to a DNP (e.g. "only" = "nothing other than") will be dotted underlined. Notice that the word "must" is a modal word, which, as all modal words, is equivalent (*via* the S4 model of modal logic) to a doubly negated proposition of intuitionist logic (a DNP); hence it is not equivalent to the corresponding affirmative proposition, owing to the failure of the double negation law of such a kind of logic.

[16] According to Heisenberg: "Methodologically, his staring points were particular problems not the wider relationship. When he described his approach, I often had the feeling that he looked upon scientific research much as some mountaineers look upon a tough climb. All that matters is to get over the next three yards. lf you do that long enough, you are bound to reach the top." (quoted in Kragh 1990, p. 281). This analogy seems to allude to a step-by-step development of a PO theory. which proceeds through local, industive steps towards the discovery of a new method.

In comparison with the frequency of DNPs within the original texts of other physical theories, the number of thirteen DNPs inside two pages of this text is high. Although one may doubt whether some DNPs play essential roles in Dirac's illustration, it is easy to recognize that the DNPs 4-11 play a decisive role for illustrating the theory.

Hence, also these DNPs give evidence for a PO organization of this theory.

## 7. The problem-based organization of Dirac's theory: the *ad absurdum* arguments

After having accumulated all mathematical tools of his theory Dirac's goal is to solve the theoretical *problem*: "Is this analogy valid?". He has to answer by exhibiting specific arguments on the subject.

The more relevant corroboration of the PO nature of a theory is the occurrence of *ad absurdum* arguments (AAAs). Dirac's text presents two such arguments; they are constituted by the DNPs no.s 4, 5 and 6 (which for reader's convenience I reiterate):

4) On the other hand, the assumptions of the four preceding chapters are so closely interconnected that one could hardly modify them in any way…. (I, p. 92)

5) … without getting an entirely different scheme of mechanics,… (I, p. 92)

6) … and the successes of the theory are so great as to make it fairly certain that no such modifications will be required.(I, p. 92)

Truly, the nature of these AAAs is a little unclear, owing to his use of allusive words. In DNP 4 he writes "one could hardly…" in place of the usual "it is absurd…" Moreover, he makes use of a singular kind of absurdity; its role is played by the words "an entirely different scheme", i.e. the failure of his entire theoretical framework. Actually, this AAA is incorrect; the absurdity cannot be obtained by a mere negation of the assumed hypothesis to be excluded, but by an argument which here is lacking.

In addition also DNP no. 6 is an AAA. The role of an absurdity is again played by the same previous words, albeit little changed (more accurately, he would have had to write: if such modifications were possible, its so great number of successes would be absurd); moreover, instead of stating an absurdity of the assumed hypothesis, he changes the above DNP into the corresponding affirmative proposition ("fairly certain"), yet he nuances its content by means of the modal word "fairly". In sum, these AAAs are incorrect. However, it is clear that in order to support his analogy Dirac feels as necessary to make recourse to AAAs.

Then he studies the algebraic properties of PB of the Hamiltonian. The set of its properties (anti-symmetry, linearity, Leibniz' rule, Jacobi's rule) is then attributed to a structure named "quantum P.B." having the form of quantum commutators. He then suggests an argument in order to prove that these properties "are sufficient to determine the form of the quantum P.B. uniquely".[17] Dirac conclude that they are similar so that he links the two objects, PB and commutators of a quantum system, by means of a proportional relation.(I, pp. 94-96) The conclusion is supported by one more AAA, located at the beginning of sect. no. 33:

> The assumption that the P.B. defined by (10) is the analogue of the classical one enables us to take over the classical equations of motion […] onto the quantum theory […] We have thus solved the problem of obtaining equations of motion and quantum conditions forming a natural generalization of the classical theory. *The classical theory is in fact, given by the limiting case of h = 0 of the quantum theory.*(I, p. 96)

In order to make more apparent the nature of Dirac argument as an AAA I translate it in the following way. The result is a natural generalization because otherwise the limiting case $h = 0$ does not hold true, that is is an absurdity. That is, the argument goes as follows: if my analogy was wrong, the limit h → 0 for QM → CM would fail; that is absurd. The last proposition constitutes a more adequate argument because the absurdity is referred to the several successful cases, not an unproved certainty ("is"), as Dirac does.

---

[17] Actually, this step of Dirac's theory is insufficiently proved. Subsequently, a great debate tried to correct this result.(Ali and Englis 2005) Eventually, Morchio and Strocchi offered an accurate solution inside a suitable Poisson C*-algebra. (Strocchi 2018, chp. 7)

As a matter of fact, after his argument Dirac makes use of the equation (10) as an axiom for the following theoretical development.

In next sect. no. 34 he starts to represent the most known subject to quantum theorists, Schroedinger's equation. First, he introduces "Schroedinger's form of quantum condition", i.e. the relation $p_x = ih/2\pi dx$. He concludes:

> This equation is [by itself] quite a plausible assumption for one to make for one's quantum conditions, [-] apart from the fact that it is derivable from equations (12) [= commutation relations of the conjugate variables], which were set up from analogy with the classical theory [-], on account of its simplicity and generality and the fact that it leads at once to the law of conservation of momentum. (I p. 110)

Dirac correctly qualifies the result of the previous argumentations as no more than "plausible". This adjective means not other than "It is not false that…", i.e. a DNP; which is the correct conclusion of a chain of arguments relying on DNPs and obtaining an analogy.

Hence, he concludes that its analogy is plausible for reasons of "simplicity and generality". These words actually constitute a (unware) application of a weak form of the principle of sufficient reason, which attributes to the relation between our minds and the reality an intimate correspondence satisfying "simplicity and generality". By scrutinizing the entire theoretical development one verifies that this logical principle plays the governing role for the entire theory, which depends on an analogy. (However, Dirac's application of it is slightly misdirected, because his words "simplicity and generality" are referred to the result, the equation, rather than the process of translation from the assumption into an axiom)

Thus, the above step, actually the final step of a PO theory (i.e. an implicit application of the principle of sufficient reason) (Drago 2012), changes the conclusive, universal predicate, which is a DNP, into its corresponding affirmative predicate. As a matter of fact, after the above quoted period Dirac proceeds by assuming as an assured axiom exactly the affirmative proposition $p_x = ih/2\pi dx$ corresponding to previous DNP. Thereafter he is capable of solving a great number of quantum dynamical problems by a deductive path from this equation considered by him as an axiom.

An additional evidence for attributin to his theory the nature of a PO comes from his particular mathematical technique, that, according to him, QM has introduced in theoretical physics. He characterizes CM theoretical representation of the reality as a space-time-force one. He calls it as the "coordinates method". Instead the new theoretical physics

> requires the use of the mathematics of transformations. The important things in the world appear as the invariants… of these transformations… (I, p. v)

Dirac refers to the canonical transformations; however, one has to intend also group theory in general, since his "transformation theory" includes the group of special relativity. He adds that this mathematical method is more difficult to be learnt, because it is more abstract then the method based on direct sensations and experiments performed within space and time. However, he declares to prefer the former one because it has to be considered "the essence of the new method in [the entire] theoretical physics".(I, p. v). Since a previous paper showed that within theoretical physics group theory is applied by theories relying on the choice PO,( Drago 1996) Dirac' belief in "transformation theory" gives further evidence for characterizing the organization of his theory as a PO.[18]

---

[18] Actually, the latter method is an ancient one. It characterizes Leibniz's approach to theoretical physics ("our minds looks for invariants"). It was reiterated in 1783 by L. Carnot ("to look for the invariants of bodies collision") in his original foundation of mechanics.(Carnot L. 1783, pp. 18 and 43). Unfortunately, Dirac ignores these anticipations and moreover he recalls (I, p. vi) Weyl's book (Weyl 1928) not because for the first time it had introduced into QM group theory, but as an unqualified exception to the coordinates method. Which Dirac does not calls "group theory" owing to his under-evaluation of it; rather he prefers to baptize it as the "symbolic" method of the invariants (I, p. vi). It is apparent from a general analysis of his use of group theory that Dirac emphasizes specific aspects of group theory, which after the full introduction of group theory within theoretical physics will be considered as only particular features.

## 8. The "Preface" of the book: the great role played by mathematics in Dirac's mind

Let us now come back to analyze the first pages of the book. The "Preface" of the first edition has been reiterated in all next editions. Within one page and half Dirac declares in a clear-cut way some important tenets and choices.

He starts by stressing the great novelty of the new theory QM, with respect to previous TP.. The new theory is

> built up from physical concepts which cannot be explained in terms of things previously known to the student, which cannot even be explained adequately in words at all. (I, p. v)

Moreover, the mathematical laws of the new theory obstruct any "mental picture" in terms of space-time-force, i.e. the background of previous theories. Rather, they allow to "control a substratum of which we cannot form a mental picture without introducing irrelevancies";(I, p. v) Therefore, the proper way of presenting QM "must necessarily be abstract".(I, p. vi) Dirac's approach is to rely his theory as much as possible on mathematics, the proper way of his presentation of QM is through abstract mathematics ("the symbolic method").

Dirac however recognizes that in QM mathematics is under debate. The "mathematical form" of QM has split in two "methods": that of "the invariants… [of] transformations" (within Matrix Mechanics) and that of "coordinates or representations" (within Wave Mechanics). He recognizes that most physicists prefer the latter one for valid reasons (historical continuity and familiarity with its formalism). But the former one (to which he recall that Weyl's book(1928) belongs)

> seems to go more deeply into the nature of things…[, it] express[es] the physical laws in a neat and concise ways and will be increasingly used in the future. For this reason I have chosen the [transformations, while] introducing the [wave mechanics] later merely as an aid to practical calculation.(I, pp. vi-vii)

His mathematics presents a very important novelty; it is that of "transformations", leading to find out the "invariants" of the physical system. Dirac states that the "essence of the new method in TP" is the use of this kind of mathematics, so that the basic program of TP consists in widening even more the kinds of transformations. (By passing, he mentions two related subjects; he stresses the introduction of an observer within TP and the greater difficulty of the didactics of the new TP.) He then declares what may be considered as a belief in mathematics: "there is no limit to its power in this field". The subject was illustrated by Pais (2005, pp. 33-36) through a list of quotations ranging along Dirac's entire professional life. One can resume his attitude through the slogan: First come all possible mathematical developments, then one sees whether they have a physical significance. Therefore it is not surprising that Dirac writes in the "Preface": "a book on the new [unforeseeable] physics…. has to be essentially mathematical."(I, p. vi) [19]

Above Dirac's general remarks manifest the fact that, in order to give the most assured theoretical basis to a physical theory, he decisively strengthened the mathematical aspects. In fact, in the Chapters following the first one he formulates the mathematical part of his theory so brilliantly that most readers of his book have been impressed mainly by it. (However, in order to not send away the readers he declares to have mitigated his mathematical formalism by supporting it as much as possible by means of experimental facts and physical explications).

## 9. Appraisal on Dirac's theoretical attitude within book's first edition

Here we will consider Dirac's revolutionary theory from the viewpoint of almost a century of subsequent QM developments. No surprise if after so much time this theory presents some shortages.

---

[19] About Dirac's conception of mathematics, Kragh (1990, chp. 14) discussed what may be called Dirac's "principle of mathematical beauty". Surely, Dirac supported this idea so much to attribute to it a decisive role. But this principle is not a well-defined subject in theoretical physics and in philosophy of science either. No surprise if Kragh's discussion did not achieve certain results.

Although the book is free of any current interpretation of QM (e.g. Copenhagen's), Dirac does interpret QM in philosophical terms. First, he gives more relevance to the superposition principle than the indeterminacy principle, whereas other physicists invert the order of derivation.

Moreover, let us consider his mathematics. By resuming his "Preface" devoted to his use of mathematics, I list his basic points:

*i*) QM is an essentially new theory;
*ii*) mathematics is a so important component of theoretical physics to assure its progressive improvement;
*iii*) between the two kinds of mathematical descriptions of the new theory, he decisively chooses the algebraic one;
*iv*) to which pertains the new technique of the "invariants… of transformations" which he characterizes as the mathematics of QM.

These points are presented without hesitation by short and incisive words, although two points of them (*iii* and *iv*) are shared by few theoretical physicists of his time. Here Dirac is a determinate theoretical physicist of his time. This Dirac's attitude differs very much from the embarrassed justifications of the organization of his theory at the end of Chapter I.

He attributed so much importance to his mathematics that one may say that his theory represents an instance of a "rational formulation of a physical theory", i.e. a theory sharing the approach of rational mechanics, which is developed on the basis of some mathematical notions summarizing physical features. Whereas "rational mechanics" relies directly on the mathematical properties of points, infinitesimals and analytical functions, his theory relies directly on vectors within an infinite vector space (where groups of transformations play a crucial role) and the algebra of commutators. Each of these theories is based on a mathematical notion from which to build the entire theory; while the former one makes use of the analytical mathematics, the latter theory the algebraic mathematics.

It is also important to remark that his mathematics is clearly based on actual infinity (AI). This fact is manifested by e.g. his introduction of $\delta(x)$ function, (I, sect. 22) which does not correspond to a constructive function. But it is also manifested, by opposition, by what he affirms in pag. 1:

> The most striking (*although not its most important*) differences from the old mechanics apparently show a discontinuity in certain physical processes and a discreteness in certain dynamical variables.(I, p.1; emphasis added)

Here it is apparent that Dirac does not give mathematical relevance to both quanta and discreteness of some formulas of QM; he considers them as particular cases which are well-represented by the general mathematical framework of QM. It was not so in Einstein's 1905 paper, whose introductory section stressed a "dichotomy" between "continuum" mathematics and "discrete" mathematics within TP (Einstein 1905).

However, subsequent history of the mathematical improvements of QM did not occurred along Dirac's suggestions. His emphatic words "symbolic method" was forgotten. The same occurred for the trick of naming a vector and its dual as bra and ket. Worst, his idea of distinguishing through two specific algebras *c*-numbers and *q*-numbers (apparently, classical and quantum numbers) failed (Darrigol 1990, p. 26) His version of $\delta(x)$ function was so unsatisfactory that it stimulated scholars to introduce both the theory of distributions and the theory of rigged Hilbert space in order to supersede it. It therefore was a good idea, but to be reformulated in more accurate, mathematical terms.

Dirac also choose the mathematical theories, i.e. algebra and group theory, which in his time were alternative to the typical AI mathematics, infinitesimal analysis and differential equations. Through his algebra he successfully constructed the theory. Two decades later, his algebra was generalized into a C*-algebra, where the notion of state-vector is derived as a consequence. (Segal 1947) But this improvement originated from the mathematics of Hilbert space.

After Dirac's book, his "theory of transformations" (actually, the technique of group transformations) did not remained a specific part of QM's mathematics– as Dirac held -; instead it gained ever more importance in theoretical physics. Applied to all physical theories, it gained the same relevance of differential equations. He therefore lost a chance for improving the basic mathematics of theoretical physics.

Hence, Dirac's great trust in his capability of advancing QM's mathematics was not supported by the subsequent development; hence its "elegance" was not an insurance of its validity or profundity, but a contingent aspect.

Over all, after von Neumann's book (1932) his mathematical approach to QM appeared a less powerful one than Hilbert space', because his mathematical formalism may be considered a reduced one with respect to von Neumann's. Moreover it was essentially inaccurate, owing to the introduction of his celebrated $\delta(x)$ function. About these points the following remarks are important:

> In the preface to his book on the foundations quantum mechanics, von Neumann says of Dirac's own formulation of quantum theory that it is "scarcely to be surpassed in brevity and elegance," but that it "in no way satisfies the requirements of mathematical rigour." […] Since [von Neumann's] work was published, little has changed to affect the validity of these remarks [….] the Dirac formalism remains far from rigorous, and the formulation in terms of Hilbert space is still the only adequate framework for quantum theory. The very elegance and success of the Dirac formalism have ensured its survival. Most of the current generation of books on quantum theory prefer to take it as their guide, rather than give more than a passing reference to the niceties of Hilbert space. The most unsatisfactory feature of the present situation is that the gulf between the Dirac formalism and Hilbert space is quite substantial, so that a lot of rethinking is necessary [to a student] before grasping the "correct" way of expressing things in Hilbert space. (Roberts 1966, p. 1097)[20]

One more shortage is that often, in order to corroborate his results, Dirac referred to the great use theoretical physicists of previous decades made of the limit $h \to 0$. In the above we saw a crucial AAA, the third one, relying on it. But recent studies stressed that this limit is singular; hence, a full reduction of QM into CM is obstructed. (Rohrlich 1990, sect. 6; Bokulich 2008, chp.s 1, 5) and *viceversa* a generalization from CM to QM is at all insecure.

All in all, Dirac's emphasis on his mathematics was well-posed with regard to his general algebraic approach to QM. But it was ill-posed with respect to many specific aspects, which the subsequent mathematical improvements of QM did not recognized as important. Rather, he did not gave relevance to a very promising mathematical novelty, group theory.

Let us now resuming his choice for PO.

The above sections 5-7 showed that book's first edition qualifies Dirac as belonging to, rather than the dominant group of theoretical physicists led by von Neumann, the few theoretical physicists supporting an alternative organization of QM. Our analysis of the organization of his theory showed that it was close to the model of a PO. This was a very advanced result because Dirac lived in a time of a general ignorance of an alternative organization to the axiomatic method. Moreover, his almost PO theory is important also because in the history of QM does not exist a formulation of QM more closely approximating the model of PO; for instance, although aimed to solve a clear problem (the spectral lines), Heisenberg's theory is far to be organized as a PO.[21] The

---

[20] Roberts offered also a detailed appraisal of Dirac's mathematical presentation of QM: "Conclusions. The Dirac formalism may be regarded as being valid for a wide range of quantum systems provided we make a number of modifications, the most important of which are: (1) The bras are in 1-1 correspondence with a subset of the kets and not with all the kets. This is already implicit in Dirac's work, because he relaxes the requirement that the complete bracket expression should always be defined. (2) The observables used in representation theory should be continuous. (3) The term "commuting observable" is to be understood in the usual Hilbert space sense of commuting spectral resolutions. (4) A d-function normalization of a continuous spectrum is only possible when the corresponding spectral measure is absolutely continuous with respect to Lebesgue measure. (5) An eigenket of an observable is only of direct physical significance if it forms part of the integral eigendecomposition, associated with the spectral resolution of the corresponding self-adjoint operator on Hilbert space." (Roberts 1966, p. 1103).

[21] Alisa Bokulich (2008, chp.s 2-4) tried to classify the differet kinds of organizations of a theory. Unfortunately she makes use of philosophical, loose distinctions. According to Bokulich, Heisenberg's theory was "close"; this is a

first attempt of founding QM upon group, i.e. Hermann Weyl's book (1928) accumulates chapters concerning very important achievements, yet without a systematic organization. Only the organization of Einstein's 1905 paper on quanta (apparently based on an analogy) is a clear PO (Drago 2013); but this paper presents not an entire formulation of QM, but a mere starting point of the history of the construction process of QM.

In sum, in the first edition of his book Dirac founded his theory upon the choices AI and PO. These are the same of Lagrange's theory, the best representative of all the theories relying on the same couple of choices.(Drago 2009). It is well-known that Lagrange's theory was celebrated for its "elegance"; also Dirac's theory was celebrated for its "elegance" (see eg previous von Neumann's quotation). Hence, Lagrange's mechanics and Dirac's theory are similar in both structural terms (i.e. the same fundamental choices) and in subjective appearances (i.e. the elegance of mathematical framework).

**10. Comparison with the next editions of the book**
The sequence of the next editions of Dirac's book shows that he not only improved the physical contents of the book, but also abandoned his uncertainty upon the kind of organization. Maybe he felt himself in malaise for his attempt to define in the first edition of his book his own alternative organization; as a fact, *he then declared to assume the previously devalued, axiomatic organization.*

With respect to the first edition of Dirac's book the second edition is somewhat different. He reiterated the "Preface" except for the last period of five rows (concerning the second half of the book, which is different from that of the first edition). He added a chapter on the quantization of electrodynamics' field and for the first time (!) a table of contents.

He changed also the Chapter I. In the first section ("General Remarks") he added to the description of a polarization experiment that of an interference; which is useful for didactical purposes; also in each point of the interference pattern again there are two values. His aim was to give a less narrow physical basis to his subsequent huge mathematics. Moreover, sect. 5 introduces a "mathematical formulation of the principle" and moreover a sect 6 introduces Dirac's mathematical names for a vector and its dual, bra and ket; the sect. 4 of the first edition ("Compatibility of the observations") is suppressed. The last two sections could be better located at the beginning of the next Chapter concerning all the notions of the vector space).

*But his warning on the kind of organization of a theory – in the first edition located at the end of the chapter 1 - is now lacking.*

Rather, he introduces a distinction between part (1) and part (ii) of the theory according to the time is fixed or variable.(II, p. 17) Actually, this novelty may be considered of a minor importance, because may be considered as the usual distinction between static and dynamics. Yet, this change conforms Dirac's theory to the usually deductive theories, whose theoretical development starts from static and then presents dynamics; this development is the contrary one of that of the first edition, starting from the dynamics of the P.B. of an Hamiltonian. Hence, this change is related to his new kind of organization; which *he declares in open words* in sect. 5, before the introduction of his mathematical notions.

> Quantum mechanics provides a good example of the new ideas. It requires the states of a dynamical system and the dynamical variables to be interconnected in quite strange ways that are unintelligible from the classical standpoint. The states and dynamical variables have to be represented by mathematical quantities of different natures from those ordinarily used in physics. The new scheme becomes a precise physical theory when all the axioms and rules of manipulation governing the mathematical quantities are specified and when in addition certain laws are laid down connecting physical facts with the mathematical formalism, so that from any given physical conditions equations between the mathematical quantities may be inferred and vice versa. In an application of the theory one would be given certain physical information, which one would proceed to express by equations

---

characteristic feature of an axiomatic theory, i.e. an AO; instead, the first formulation of QM clearly lacked of axioms and rather was based on a problem, hence a PO. In op position, she characterizes Dirac's theory as "open" with respect to next developments, not through its structural features of present times.

between the mathematical quantities. One would then deduce new equations with the help of the axioms and rules of manipulation and would conclude by interpreting these new equations as physical conditions. The justification for the whole scheme depends, apart from internal consistency, on the agreement of the final results with experiment. (IV p. 14-15) [22]

Maybe after the first edition the success of his mathematics led Dirac to disregard in the second edition the "philosophical" question of the organization of his theory, for rather accepting the rigorously deductive-axiomatic attitude of the mathematical apparatus on which he trustfully relied.[23]

In the subsequent editions he introduced more changes and further new improvements. In the fourth edition of the year 1958 – whose contents are the final ones - also the first chapters are changed; in particular the number of the theoretical chapters is 5 instead of 6. However, the essential contents of the first section of the chapter VI of the first edition are reiterated by those of the first section (no. 25) of the chp. V of the second edition and also by those of the first section (no. 21) of the Chp. IV of the fourth edition. Yet, his discussion is shortened from two pages to half a page; all above lsted DNPs have been cut out (apart the word "analogy", which alone preserves the inductive nature of Dirac's theory).

**11. Conclusions**
Dirac's first edition of the celebrated book was organized according to the model not of an axiomatic theory, in which he declared to distrust, but a PO. He applied this model of organization to his theory in almost a complete way. However, since the second edition (in the year 1935) he abandoned it for applying instead the previously rejected model of axiomatic organization. This model, together with his sophisticated use of mathematics has been commonly interpreted as a mere variant of the subsequent von Neumann's axiomatic theory, so much that current QM is commonly called "Dirac-von Neumann" formulation.

But Roberts warned that there remained a "gulf" between the two formulations, owing to many technical reasons. It is enough to stress that Dirac's approach is an algebraic one, whereas von Neumann's is an analytical one. Moreover, it is commonly ignored that just in the same year of second edition of his book, also the latter one changed attitude with respect QM: in a letter to Garrett Birkhoff he expressed a dramatic "confession", he distrust in Hilbert space since this mathematical notion is inadequate for faithfully representing QM:

> I would like to make a confession which may seem immoral: I do not believe absolutely in Hilbert space any more. After all, Hilbert space (as far as quantum mechanical things are concerned) was obtained by generalizing Euclidean space, footing on the principle of 'conserving the validity of all formal rules' [...]. Now we begin to believe that it is not the vectors which matter, but the lattice of all linear (closed) subspaces. Because: 1) The vectors ought to represent the physical states, but they do it redundantly, up to a complex factor, only 2) and besides, the states are merely a derived notion, the primitive (phenomenologically given) notion being the qualities which correspond to the linear closed subspaces.(von Neumann 2005)

Later, he embraced algebra. His next researches on the foundations of QM wanted to find out new *algebraic* structures joining logic and probability, but unsuccessfully (apart his discovery, together with Birkhoff, of the non-classical logic of QM in the year 1936)

In sum, present formulation of QM born in the year 1930 through the PO of the first edition of Dirac's book; but this organization was quickly abandoned by Dirac. After him, in the 1932 von Neumann founded anew QM. But in the same year (1935) of Dirac's dismissing PO, he no longer believed that the basic mathematical notion of his formulation, Hilbert space, was adequate to represent QM. Hence, this almost infallible physical theory in its results, QM, born according to two lines of research, Dirac's and von Neumann's, which have been denied by their two respective founders.

---
[22] To my knowledge no contemporary physicist remarked a change in Dirac's attitude on the kind of organization, nor his change of the basic philosophy occurred in the fourth edition.
[23] Notice that in the same years Emmy Noether changed algebra according to an axiomatic attitude in a so radical way that H. Weyl felt himself as an outdated matematician (Weyl 1936, "Preface").

Let us add that QM was formulated by them before the discovery of quantum logic (1936), which would require a way of reasoning on the foundations of the theory radically different from the classical one. Dirac did not tried to re-formulate QM according to this novelty and von Neumann was unsuccessful. Hence the formulation of QM remained the usual one, which is based on the classical logic, although we know that it is different from the appropriate one.

In sum, both present Dirac's and von Neumann's formulations of QM lead to theoretical schemes fortunately obtaining valid experimental results; but at present time these formulations appear without a certain foundation because they do not answer to the basic new dichotomies discovered by themselves: i) Birkhoff-von Neumann's dichotomy about logic (classical, or non-classical?), *ii*) Dirac's dichotomy about the model of organization (axiomatic or problem-based?), *iii*) Dirac's and von Neumann's dichotomy about the kind of mathematics (analytical or algebraic?). These are the true "hidden variables" of the foundations of QM. It is with respect to them that this theory has to achieve more clear foundations.

**References**

Ali S.T., Englis M. (2005). "Quantization Methods: A Guide for Physicists and Analysts". *Review of Mathematical Physics*, 17, pp. 391-490.
Bokulich A. (2008), *Reexamining the Quantum-Classical Relation, Beyond Reductionism and Pluralism*, Cambridge: Cambridge U.P.
Brown L.M. (2006), "Paul A.M. Dirac's *The Principles of Quantum Mechanics*", *Perspectives in Physics*, 8, pp. 381-407.
Carnot L. (1783), *Essai sur les Machines en général*, Dijon: Defay.
Carnot L. (1803), *Principes fondamentaux de l'équilibre et du mouvement*. Deterville, Paris.
D'Agostino S. (2002), "From Rational Numbers to Dirac's Bra and Ket: Symbolic Representation of Physical Laws", *Physical Perspectives*, 4, 216-229.
Darrigol O. (1990), "Dirac P.A.M.", in F.L. Holmes (ed.), *Dictionary of Scientific Biography*, New York: Charles Scribner's Sons, Vol. 17, Supplement II, pp. 224-233.
Dirac P.A.M. (1930), *Principles of Quantum Mechanics*. Oxford: Oxford U.P..
Drago: A. (1996), "Una caratterizzazione del contrasto tra simmetrie ed equazioni differenziali", in Rossi A. (ed.): *Atti XIV e XV Congr. Naz. St. Fisica*, Lecce: Conte, pp. 15-25.
Drago A, (2007), "There exist two models of organization of a scientific theory", *Atti della Fond. G. Ronchi*, 62 n. 6, pp. 839-856.
Drago A, (2009), "The Lagrange's arguing in *Méchanique Analytique*", in Sacchi Landriani G. and Giorgilli A. (eds.): *Sfogliando la Méchanique Analvtique. Giornata di Studio su Louis Lagrange,* Milano: LED, pp. 193-214.
Drago A. (2012), "The Relationship Between Physics and Mathematics in the XlXth Century: The Disregarded Birth of a Foundational Pluralism", in E. Barbin and R. Pisano (eds.): *The Dialectic Relations Between Physics and Mathematics in.the XIXth Century*, Berlin: Springer, 2013, pp. 159-179.
Drago A. (2013), "The emergence of two options from Einstein's first paper on Quanta", in R. Pisano, D. Capecchi, A. Lukesova (eds.), *Physics, Astronomy and Engineering. Critical Problems in the History of Science and Society*, Siauliai: Scientia Socialis P., pp. 227-234.
Drago A., Manno S.D. (1989) "Le ipotesi fondamentali della meccanica secondo Lazare Carnot". *Epistemologia* 12, pp. 305–330.
Dugas R. (1950,) *Histoire de la Mécanique*. Neuchâtel: Griffon.
Einsten A. (1905), .Ueber einen die Erzeugung der Verwandlung des Lichtes betreffenden heuristisch Gesichtpunkt., *Ann. der Physik*, 17, pp. 132-148; reprinted in Stachel J. (ed.) (1989), *Collected Papers of Albert Einstein*, Princeton, Princeton U.P., vol. 2, 149-165.
Frisch M. (2006). "Mechanics, principles, and Lorentz' cautious realism". *Studies in History and Philosophy of Modern Physics*, 36, pp. 659-679)
Gutzwiller M.C. (1998), "Resource Letter ICQM-1: The Interplay between Classical and Quantum Mechanics", *American Journal of Physics*, 66,(4), pp. 304-319.

Jammer D. (1974), *The Philosophical Development of Quantum Mechanics*, New York: Wiley, p. 389.
Jammer D. (1989), (1966). *Conceptual History of Quantum Mechanics*. New York: Mc Graw-Hill.
Kragh H. (1990), *Dirac: A Scientific Biography*, Cambridge: Cambridge U.P. 1990.
Kragh H. (2013), "Paul Dirac and The Principles of Quantum Mechanics", in Bandino M. and Navarro J., *Research and Pedagogy. The History of Quantum Physics through the Textbooks*, Berlin: Ed. Open Access, pp. 249-264; http://edition-open-access.de/studies/2/index.html.
Koopman B.O. (1935), "Dirac on quantum mechanics", *Bull. Am. Math. Soc.*, 41, pp. 471-474.
F. Kronz and T. Lupher (2019) "Quantum Theory and Mathematical Rigor", in N.E. Zalta (ed.), *Stanford Encyclopedia of Philosophy*. https://plato.stanford.edu/entries/qt-nvd/
Lagrange J.-L- (1788), *MécaniqueAnalytique*, Paris : Desaint.
Mehra J., Rechenberg H. (1982), *The Historical Development of Quantum Theory*, Vol. 2: The Discovery of Quantum Mechanics, Berlin: Springer
Pais A.n (2005), "Paul Dirac Aspects of his life and work", in P. Goddard (ed.); *Paul Dirac. The man and His Work*, Cambridge: Cambridge U.P., pp. 1-45.
Pauli W. (1932), "Letter to Ehrenfest", October 28.
Poincaré, H. (1903), *La Science et l'Hypothèse*, Paris : Hermann.
Poincaré, H. (1905), *La Valeur de la Science*, Paris : Flammarion.
Roberts J. E. (1966), "The Dirac Bra and Ket Formalism", *Journal of Mathematical Physics*, 7, no. 6, pp. 1097-1104.
Rorhlich F. (1990) "There is good physics in te reduction", *Foundations of Physics*, 20, no. 11, pp. 1399-1411.
Segal I. (1947), "Postulates of General Quantum Mechanics". *Annals of Mathematics*, 48, pp. 930-948. https://doi.org/10.2307/1969387
Strocchi F. (2018). *A Primer of Analytical Mechanics*, Berlin: Springer.
von Neumann J. (1932), *Grundlagen der Quantum Mechanik*, Berlin: Springer.
von Neumann J. (2005), *Selected Letters*, M. Rédei (ed.), Providence: American Mathematical Society.
Weyl H. (1928), *Group Theory and Quantum Mechanics*, New York: Dover.
Weyl H. (1936), *Cassical Groups* ???
Wigner E. (1959), "Preface", to *Group theory and Application to Quantum Mechanics and Atomic Spectra* (1931), Acad. P., New York.